\documentclass[10pt,a4paper]{article}

\usepackage[utf8]{inputenc}
\usepackage[english]{babel}

\begin{document}
\large

\newpage
\begin{center}
\LARGE{\bf An Unbroken Axial-Vector Current 
\\Conservation Law}
\end{center}
\vspace{0.1mm}
\begin{center}
{\bf Rasulkhozha S. Sharafiddinov}
\end{center}
\vspace{0.1mm}
\begin{center}
{\bf Institute of Nuclear Physics, Uzbekistan Academy of Sciences,
\\Tashkent, 100214 Ulugbek, Uzbekistan}
\end{center}
\vspace{0.1mm}

\begin{center}
{\bf Abstract}
\end{center}

The mass, energy and momentum of the neutrino of a true flavor have an axial-vector nature. 
As a consequence, the left-handed truly neutral neutrino in an axial-vector field of emission 
can be converted into a right-handed one and vice versa. This predicts the unidenticality 
of masses, energies and momenta of neutrinos of the different components. Recognizing such a difference in masses, energies, momenta and accepting that the left-handed axial-vector neutrino 
and the right-handed antineutrino of true neutrality refer to long-lived C-odd leptons, and the right-handed truly neutral neutrino and the left-handed axial-vector antineutrino are of short-lived fermions of C-oddity, we would write a new CP-even Dirac equation taking into account the flavor symmetrical axial-vector mass, energy and momentum matrices. Their presence explains the spontaneous mirror symmetry violation, confirming that an axial-vector current conservation law has never violated. They reflect the availability of a mirror Minkowski space in which a neutrino is characterized by left as well as by right space-time coordinates. Therefore, it is not surprising that whatever the main purposes experiments about a quasielastic axial-vector mass say in favor of 
an axial-vector mirror Minkowski space-time. 

\vspace{0.8cm}
\noindent
{\bf 1 Introduction}
\vspace{0.4cm}

There is a range of fundamental differences [1] in nature of currents of vector and axial-vector 
types, which suggest the classification of elementary particles with respect to C-operation [2]. 
This procedure in turn admits the existence of leptons of true $(l^{A}=C{\bar l^{A}})$ neutrality. Any of these forms of leptons must have his own Dirac $(\nu_{l}^{A}=C{\bar \nu_{l}^{A}})$ neutrino. They are united in families of fermions of an axial-vector $(A)$ nature. 

It is not excluded, however, that to each type of a C-odd Dirac $\nu_{l}^{A}({\bar \nu_{l}^{A}})$
neutrino corresponds a kind of Majorana $\nu_{M}^{A}({\bar \nu_{M}^{A}})$ neutrino [3]. Such pairs 
constitute the purely neutrino families [4].

A truly neutrality of a neutrino of this type implies the absence in such a fermion of a vector 
C-invariant electric charge and of all earlier known [5,6] lepton flavors. Its distinction from 
his antiparticle becomes possible $(\nu^{A}=-CP{\bar \nu^{A}})$ owing to the CP-symmetrical
behavior [2] of their axial-vector C-noninvariant electric charge [1]. Therefore, the nature 
itself distinguishes any family of these types of fermions from others by a kind of true flavor. 

Such a new flavor symmetry suggested by the author [2] requires one to characterize each C-odd particle by the three $(l^{A}=e^{A},$ $\mu^{A},$ $\tau^{A})$ true flavors:
\begin{equation}
T_{l^{A}}=\left\{
{\begin{array}{l}
{-1\quad \mbox{for}\quad l^{A}_{L}, \, \, \, \, \, l^{A}_{R}, \, \, \, \,
\nu_{lL}^{A}, \, \, \, \, \nu_{lR}^{A}, \, \, \, \,
\nu_{ML}^{A}, \, \, \, \, \nu_{MR}^{A},}\\
{+1\quad \mbox{for}\quad \overline{l}^{A}_{R}, \, \, \, \,
\overline{l}^{A}_{L}, \, \, \, \,
{\bar \nu_{lR}^{A}}, \, \, \, \, {\bar \nu_{lL}^{A}}, \, \, \, \,
{\bar \nu_{MR}^{A}}, \, \, \, \, {\bar \nu_{ML}^{A}},}\\
{\, \, \, \, \, 0\quad \mbox{for}\quad \mbox{remaining particles.}}\\
\end{array}}\right.
\label{1}
\end{equation}

One of the most highlighted features of conservation of full true number 
\begin{equation}
T_{e^{A}}+T_{\mu^{A}}+T_{\tau^{A}}=const
\label{2}
\end{equation}
or all forms of true flavors
\begin{equation}
T_{l^{A}}=const
\label{3}
\end{equation}
is the connection between the two left (right)-handed fermions having an axial-vector nature. 
They may symbolically be written as
\begin{equation}
(l^{A}_{L}, \overline{l}^{A}_{R}), \, \, \, \,
(l^{A}_{R}, \overline{l}^{A}_{L}),
\label{4}
\end{equation}
\begin{equation}
(\nu_{lL}^{A}, {\bar \nu_{lR}}^{A}), \, \, \, \,
(\nu_{lR}^{A}, {\bar \nu_{lL}}^{A}),
\label{5}
\end{equation}
\begin{equation}
(\nu_{ML}^{A}, {\bar \nu_{MR}}^{A}), \, \, \, \,
(\nu_{MR}^{A}, {\bar \nu_{ML}}^{A}).
\label{6}
\end{equation}

It is already clear from them that the left (right)-handed axial-vector neutrino of any type 
in the field of emission similarly to all C-odd leptons [7] can be converted into a right 
(left)-handed one without change of his true flavor [8]. This is of course intimately connected 
with the availability in the neutrino of a self inertial mass, namely, the mass of a fermion leads 
to the flip of its spin. However, in the same form as it was developed, the unified field theory 
of elementary particles is not in a state to substantiate at a given stage the dynamical origination 
of such transitions in which a mass comes forward as a spontaneity criterion of mirror symmetry violation [4].  

According to one of predictions of this theory, an axial-vector current conservation law is 
basically violated at the expense of a nonzero rest mass, the absence of which would imply 
its compatibility with a chiral invariance. We have, thus, a highly peculiar situation when 
an inertial mass itself requires in principle one to go away from the earlier definition 
of the structure of matter fields.

Furthermore, if it turns out that in the same neutrino there are no those properties between which
would exist some paradoxical contradictions, from the point of view of a unity of symmetry laws, it should be expected that each transition of
\begin{equation}
\nu_{lL}^{A}\leftrightarrow \nu_{lR}^{A}, \, \, \, \, 
{\bar \nu}_{lR}^{A}\leftrightarrow {\bar \nu}_{lL}^{A},
\label{7}
\end{equation}
\begin{equation}
\nu_{ML}^{A}\leftrightarrow \nu_{MR}^{A}, \, \, \, \,  
{\bar \nu_{MR}}^{A}\leftrightarrow {\bar \nu_{ML}}^{A}
\label{8}
\end{equation}
can carry out in the field of emission as an indication in favor of so far unobserved unified regularity of the nature of mass, energy and momentum.

Therefore, to formulate the expected regularity and its consequences, we investigate here the 
question as to whether there exists any connection between the mass of a C-odd neutrino and its
spin and, if so, what the observed interconnection says about a mirror symmetry violation as well 
as an axial-vector current conservation. This does not exclude of course from the discussion the 
ideas of each of mass, energy and momentum in the spin polarization type dependence. Insofar as 
the fate of chiral invariance in nature is concerned, it will be illuminated in the separate work.

\vspace{0.8cm}
\noindent
{\bf 2 Incompatibility Criterion for the Classical Dirac Equation}
\vspace{0.4cm}

The equation of the unified field theory of particles with the spin $1/2$ written by Dirac [9] 
is based logically on the fact that the relationship involving the mass, energy and momentum
\begin{equation}
E^{2}={\bf p}^{2}+m^{2},
\label{9}
\end{equation}
follows from
\begin{equation}
\hat H=\alpha \cdot\hat {\bf p}+\beta m
\label{10}
\end{equation}
only if the matrices $\alpha,$ $\beta$ and $\gamma_{5}$ have in it the following structure
\begin{equation}
\alpha= {{0 \ \, \, \, \, \sigma}\choose{\sigma \, \, \, \, \ 0}}, \, \, \, \,
\beta={{I \ \, \, \, \, \ 0}\choose{0 \ -I}}, \, \, \, \,
\gamma_{5}={{0 \, \, \, \, \ I}\choose{I \ \, \, \, \, \ 0}},
\label{11}
\end{equation}
where $I$ is a unity $2\times 2$ matrix, and $\sigma$ are the Pauli spin matrices. It expresses, 
in the case of $E=i\partial_{t},$ ${\bf p}=-i\partial_{{\bf x}}$ and the neutrino with the 
four-component wave function $\psi(t, {\bf x}),$ the idea about that
\begin{equation}
i\partial_{t}\psi=\hat H\psi.
\label{12}  
\end{equation} 

It should also be chosen the matrices $\gamma^{\mu}=(\beta, \beta \alpha)$ on account of 
$\partial_{\mu}=\partial/\partial{\it x}^{\mu}=(\partial/\partial t, -\nabla),$ 
because of which the equation (\ref{12}) takes the form 
\begin{equation}
(i\gamma^{\mu}\partial_{\mu}-m)\psi=0.
\label{13}
\end{equation}

The latter would seem to convince us here that the divergence
\begin{equation}
\partial_{\mu}j^{\mu}_{A_{l}}=2im\overline{\psi}\gamma^{5}\psi
\label{14}  
\end{equation} 
of an axial-vector $A_{l}$ current
\begin{equation}
j^{\mu}_{A_{l}}=\overline{\psi}\gamma^{5}\gamma^{\mu}\psi
\label{15}  
\end{equation} 
can have the form of the continuity equation only for those fermions in which a self inertial 
mass is absent.

However, at such situations, there arises a range of nontrivial problems connected with the 
mass and charge. The point is that the mass and charge stating that the same massive particle 
must possess simultaneously both charge and current [7,8] correspond to the two forms of the same regularity of the nature of matter. Therefore, it is not surprising that if this situation follows from the mass-charge duality [10], the absence of one of mass or charge would imply that
an axial-vector current has never existed in the neutrino of true neutrality.

At the same time, it is interesting that any paraneutrino of (\ref{5}) and (\ref{6}) can also 
explain the conservation [2] of summed C-odd electric charge in those processes in which an interaction between the truly neutral fermion and the field of emission originates at the expense 
of exchange by an axial-vector photon. Its existence must be accepted as a C-invariance criterion 
for the gauge boson [11]. The Coulomb field of this type establishes in addition a broken symmetry 
of the C-noninvariant Dirac Lagrangian concerning the local axial-vector transformation [2].

If we now take into account that a classification of elementary particles and matter fields 
with respect to C-operation is fully compatible [11] with ideas of the unified field theory 
gauge invariance, then there arises a question of whether a nonzero value of the divergence (\ref{14}) is not strictly nonverisimilar even in the presence of a self inertial mass of 
a CP-even particle itself.

Here it is relevant to note that the massive neutrino can possess the longitudinal as well 
as the transversal polarizations. This does not imply of course that the same neutrino must 
be simultaneously both a longitudinal fermion and a transversal one. There exists, however, the possibility that at the availability of a nonzero mass, the longitudinal (transversal) polarized neutrino in the field of emission can be converted into a transversal (longitudinal) polarized
one without change of his flavor [12]. 

The spin transitions of these types together with ideas of mass-charge duality [10] lead us 
to an implication that a mass, by itself, does not transform the longitudinal polarization into 
a transversal one and vice versa. In other words, the mass of the neutrino can influence on its 
polarization only in the case when there are some fundamental differences in energies as well as in momenta of longitudinal and transversal neutrinos. This possibility is realized if neutrinos of the different polarizations have unidentical masses [12]. Such a difference, however, is not forbidden by a C-odd electric charge conservation law. Thereby, it requires one to restore a broken axial-vector current conservation in the unified dynamical structure dependence of all symmetry laws.

\vspace{0.8cm}
\noindent
{\bf 3 Axial-Vector Mass, Energy and Momentum Matrices}
\vspace{0.4cm}

The preceding reasoning says that the same inertial rest mass that explains in nature the 
spontaneous mirror symmetry absence, by itself, does not violate the CP-even axial-vector 
current conservation law, and the C-noninvariant neutrinos possess some so far unobserved 
universal properties. Such properties can, for example, be unidentical masses, energies and momenta of neutrinos of the different components. Recognizing this difference in masses, energies, momenta 
and accepting that the left-handed axial-vector neutrino and the right-handed antineutrino of true neutrality refer to long-lived C-odd leptons, and the right-handed truly neutral neutrino and the left-handed axial-vector antineutrino are of short-lived fermions of C-oddity, we would change 
both the mathematical logic and the compound structure of the CP-even Dirac equation in 
the presence of the flavor symmetrical axial-vector mass, energy and momentum matrices
\begin{equation}
m_{s}={{0 \ \, \, \, \, m_{A}}\choose{m_{A} \, \, \, \, \ 0}}, \, \, \, \,
E_{s}={{0 \ \, \, \, \, E_{A}}\choose{E_{A} \, \, \, \, \ 0}}, \, \, \, \,
{\bf p}_{s}={{0 \ \, \, \, \, {\bf p}_{A}}\choose{{\bf p}_{A} \, \, \, \, \ 0}},
\label{16}
\end{equation}
\begin{equation}
m_{A}={{m_{L} \, \, \, \, 0}\choose{\ 0 \, \, \, \, \ m_{R}}}, \, \, \, \,
E_{A}={{E_{L} \, \, \, \, 0}\choose{\ 0 \, \, \, \, \ E_{R}}}, \, \, \, \,
{\bf p}_{A}={{{\bf p}_{L} \, \, \, \, 0}\choose{\ 0 \, \, \, \, \ {\bf p}_{R}}},
\label{17}
\end{equation}
where $A$ comes forward as an index of the block matrix.

Unlike the neutrino of a C-even charge, the mass of which is strictly a vector $(V)$ type, particles of a C-odd electric charge have the mass of an axial-vector $(A)$ nature [13]. Therefore, it is clear that (\ref{16}) and (\ref{17}) refer to those fermions among which there are no elementary objects with vector masses, energies and momenta [14].

The account of these matrices and that a distinction of a C-odd particle and its antiparticle
\begin{equation}
\nu_{lL,R}^{A}\neq {\bar \nu_{lR,L}}^{A}, \, \, \, \,
\nu_{ML,R}^{A}\neq {\bar \nu_{MR,L}}^{A},
\label{18}
\end{equation}
as has been mentioned earlier [2], takes place at their CP-invariance, leads us to a new equation of the unified field theory of fermions of true neutrality. For the case when to a truly neutral type 
of neutrino corresponds a kind of four-component wave function $\psi_{s}(t_{s}, {\bf x}_{s}),$ it behaves as follows
\begin{equation}
i\frac{\partial}{\partial t_{s}}\psi_{s}=\hat H_{s}\psi_{s},
\label{19}  
\end{equation} 
where one must keep in mind that
\begin{equation}
\hat H_{s}=\alpha \cdot\hat {\bf p_{s}}+\beta m_{s}.
\label{20}
\end{equation}
Here $E_{s}$ and ${\bf p}_{s}$ describe the quantum energy and momentum operators 
\begin{equation}
E_{s}=i\frac{\partial}{\partial t_{s}}, \, \, \, \,  
{\bf p}_{s}=-i\frac{\partial}{\partial {\bf x}_{s}}.
\label{21}
\end{equation}

The index $s$ comes forward in them as well as in (\ref{16}) as the helicity expressing the idea of 
the unidenticality law of the space-time coordinates $(t_{s}, {\bf x}_{s})$ and lifetimes $\tau_{s}$ for the left- and right-handed neutrinos. This correspondence principle is the general and does not depend of whether a neutrino refers to vector [14] or axial-vector fermions. It is important only that the mass of the neutrino of a C-even or a C-odd electric charge was a vector or an axial-vector type. Then it is possible, for example, to use the size of $m_{s}$ in (\ref{20}) as the quantum axial-vector mass operator 
\begin{equation}
m_{s}=-i\frac{\partial}{\partial \tau_{s}}.
\label{22}
\end{equation}

If choose a particle energy $E_{s}$ and its momentum ${\bf p}_{s},$ at which 
$\partial_{\mu}^{s}=\partial/\partial{\it x}^{\mu}_{s}=(\partial/\partial t_{s}, -\nabla_{s}),$ 
then jointly with the Dirac matrices and the quantum operators 
\begin{equation}
\partial_{\mu}^{s}=
{{0 \ \, \, \, \, \partial_{\mu}^{A}}\choose{\partial_{\mu}^{A} \, \, \, \, \ 0}}, \, \, \, \,
\partial_{\mu}^{A}=
{{\partial_{\mu}^{L} \, \, \, \, 0}\choose{\ 0 \, \, \, \, \ \partial_{\mu}^{R}}},
\label{23}
\end{equation}
the equation (\ref{19}) states that
\begin{equation}
(i\gamma^{\mu}\partial_{\mu}^{s}-m_{s})\psi_{s}=0.
\label{24}
\end{equation}

From such a point of view, an axial-vector $A_{l}$ current can be leaded to his latent united form
\begin{equation}
j^{\mu}_{A_{l}}=\overline{\psi}_{s}\gamma^{5}\gamma^{\mu}\psi_{s},
\label{25}  
\end{equation} 
and its divergence in the massive case of a C-noninvariant neutrino becomes truly physical 
continuity equation
\begin{equation}
\partial_{\mu}^{s}j^{\mu}_{A_{l}}=0.
\label{26}  
\end{equation} 

So we must recognize that the difference in lifetimes of the left- and right-handed neutrinos 
of true neutrality explains the spontaneous mirror symmetry absence and the dynamical origination 
of their unidentical masses, energies and momenta responsible for conservation of an axial-vector current. Thereby, it establishes the full spin structure of the unified field theory equation of truly neutral particles with the spin $1/2$ in which a self value $s=L=-1$ $(s=R=+1)$ of the 
helicity operator $\sigma{\bf p}_{s}=s|{\bf p}_{s}|$ is definitely predicted that
\begin{equation}
\sigma{\bf p_{L}}=-|{\bf p}_{L}|, \, \, \, \, \sigma{\bf p_{R}}=|{\bf p}_{R}|.
\label{27}
\end{equation}

To show their features, we present the field of the free neutrino in the form
\begin{equation}
\psi_{s}=u_{s}({\bf p}_{s}, \sigma)e^{-ip_{s} \cdot{\it x}_{s}}, \, \, \, \, E_{s}>0.
\label{28}
\end{equation}

It is clear, however, that (\ref{11}) and (\ref{16}) separate the four-component spinor $u_{s}$
into two two-component spinors. Therefore, it should be replaced this function with
\begin{equation}
u_{s}=u^{(r)}=\left[\chi^{(r)}\atop u_{a}^{(r)}\right].
\label{29}
\end{equation}
Here $u^{(r)}$ and $u_{a}^{(r)}$ are distinguished from one another by an index $a,$ introduction 
of which is not forbidden by any conservation law.

From (\ref{19}) and (\ref{28}) one can pass herewith to a system
\begin{equation}
E_{A}u_{a}^{(r)}=(\sigma{\bf p}_{A})\chi^{(r)}+m_{A}u_{a}^{(r)},
\label{30}
\end{equation}
\begin{equation}
E_{A}\chi^{(r)}=(\sigma{\bf p}_{A})u_{a}^{(r)}-m_{A}\chi^{(r)}.
\label{31}
\end{equation}

Solving these equations for the two two-component spinors, we get  
\begin{equation}
u^{(r)}=\sqrt{E_{A}-m_{A}}
\left[\chi^{(r)}\atop \frac{(\sigma{\bf p}_{A})}{E_{A}-m_{A}}\chi^{(r)}\right].
\label{32}
\end{equation}

Taking into account that 
\begin{equation}
\chi^{(1)}=\pmatrix{1\cr 0}, \, \, \, \, \chi^{(2)}=\pmatrix{0\cr 1},
\label{33}
\end{equation}
it is not difficult to express the finding solutions in an explicit form 
\begin{equation}
u^{(1)}=\sqrt{E_{L}-m_{L}}
\left[
\begin{array}{c}
1\\ 0\\ \frac{(\sigma{\bf p}_{L})}{E_{L}-m_{L}}\\ 0
\end{array}
\right],
\label{34}
\end{equation}
\begin{equation}
u^{(2)}=\sqrt{E_{R}-m_{R}}
\left[
\begin{array}{c}
0\\ 1\\ 0\\ \frac{(\sigma{\bf p}_{R})}{E_{R}-m_{R}}
\end{array}
\right].
\label{35}
\end{equation}

Simultaneously, as is easy to see,  $u^{(1)},$ $\chi^{(1)}$ and $u_{a}^{(1)}$ correspond to the 
left-handed truly neutral neutrino, and $u^{(2)},$ $\chi^{(2)}$ and $u_{a}^{(2)}$ respond to the 
right-handed axial-vector neutrino.

In the same way one can choose the field of the free antineutrino 
\begin{equation}
\psi_{s}=\nu_{s}({\bf p}_{s}, \sigma)e^{-ip_{s} \cdot{\it x}_{s}}, \, \, \, \, E_{s}<0.
\label{36}
\end{equation}

The matrices (\ref{11}) and (\ref{16}) replace the four-component spinor $\nu_{s}$ for
\begin{equation}
\nu_{s}=\nu^{(r)}=\left[\nu_{a}^{(r)}\atop \chi^{(r)}\right],
\label{37}
\end{equation}
where $a$ must be considered as an index distinguishing $\nu^{(r)}$ and $\nu_{a}^{(r)}$
from one another.

Uniting (\ref{36}) and (\ref{37}) with (\ref{19}), one can find again that
\begin{equation}
|E_{A}|\chi^{(r)}=-(\sigma{\bf p}_{A})\nu_{a}^{(r)}-m_{A}\chi^{(r)},
\label{38}
\end{equation}
\begin{equation}
|E_{A}|\nu_{a}^{(r)}=-(\sigma{\bf p}_{A})\chi^{(r)}+m_{A}\nu_{a}^{(r)}.
\label{39}
\end{equation}

This system in turn suggests the connections 
\begin{equation}
\nu_{a}^{(r)}=\frac{-(\sigma{\bf p}_{A})}{|E_{A}|-m_{A}}\chi^{(r)}, \, \, \, \,
\chi^{(r)}=\frac{-(\sigma{\bf p}_{A})}{|E_{A}|+m_{A}}\nu_{a}^{(r)}.
\label{40}
\end{equation}

Their account leads us to the conclusion that
\begin{equation}
\nu^{(r)}=\sqrt{|E_{A}|-m_{A}}
\left[\frac{-(\sigma{\bf p}_{A})}{|E_{A}|-m_{A}}\chi^{(r)}\atop \chi^{(r)}\right].
\label{41}
\end{equation}

According to these results, the solutions of the equation (\ref{19}) for the antineutrino fields
become fully definite and behave as 
\begin{equation}
\nu^{(1)}=\sqrt{|E_{L}|-m_{L}}
\left[
\begin{array}{c}
\frac{-(\sigma{\bf p}_{L})}{|E_{L}|-m_{L}}\\ 0\\ 1\\ 0
\end{array}
\right],
\label{42}
\end{equation}
\begin{equation}
\nu^{(2)}=\sqrt{|E_{R}|-m_{R}}
\left[
\begin{array}{c}
0\\ \frac{-(\sigma{\bf p}_{R})}{|E_{R}|-m_{R}}\\ 0\\ 1
\end{array}
\right].
\label{43}
\end{equation}

We see that $\nu^{(1)},$ $\chi^{(1)}$ and $\nu_{a}^{(1)}$ characterize the right-handed axial-vector antineutrino, and $\nu^{(2)},$ $\chi^{(2)}$ and $\nu_{a}^{(2)}$ describe the left-handed truly neutral antineutrino.

Thus, the compound structure of both types of solutions (\ref{32}) and (\ref{41}) leads us to (\ref{27}) once more, confirming that the neutrino $\nu^{A}_{L}$ and the antineutrino 
${\bar \nu}^{A}_{R}$ refer to the left-polarized C-odd leptons, and the neutrino $\nu^{A}_{R}$  
and the antineutrino ${\bar \nu}^{A}_{L}$ are the right-polarized fermions of C-oddity. 

These situations and all what axial-vector mass, energy and momentum say about matter fields,
testify that our implications have the generality for any C-odd principle interacting in the framework of the equation (\ref{19}), because they follow from the unified principle.
Formulating more concretely, one can present $\psi_{s}$ as the following fields: 
\begin{equation}
\psi_{s}=\pmatrix{\psi\cr \phi}, \, \, \, \,
\psi=\pmatrix{\psi_{L}\cr \psi_{R}}, \, \, \, \,
\phi=\pmatrix{\phi_{L}\cr \phi_{R}}.
\label{44}
\end{equation}

At the unification of (\ref{44}) with (\ref{19}) there arises a united system, the solution
of which concerning $\psi_{L,R}$ and $\phi_{L,R}$ gives the right to establish one more highly important relationship
\begin{equation}
E_{s}^{2}={\bf p}_{s}^{2}+m_{s}^{2}.
\label{45}
\end{equation}

This connection describes a situation when the left-handed neutrino of true neutrality in the 
same space-time cannot be converted into a right-handed one and vice versa. Such transitions, however, appear in an axial-vector field of emission as a consequence of the ideas of 
conservation laws of C-noninvariant types of charges and currents. 

They reflect the availability of a mirror Minkowski space-time and thereby require in principle 
to characterize any C-odd neutrino by left $[(t_{L}, {\bf x}_{L})]$ as well as by right 
$[(t_{R}, {\bf x}_{R})]$ space-time coordinates. Thus, if (\ref{22}) is of fundamental laws of quantum mechanics, $\tau_{L}$ and $\tau_{R}$ must be accepted in it as the lifetimes of a truly neutral neutrino in the left and right Minkowski spaces. 

\vspace{0.8cm}
\noindent
{\bf 4 Conclusion}
\vspace{0.4cm}

From the point of view of any of a C-odd neutrino, each interconversion of (\ref{7}) and (\ref{8})
originates as the transition between the usual left (right)-handed and the mirror right (left)-handed spaces in which a neutrino of the same flavor is characterized by the different masses, energies and momenta. In other words, the particles $\nu^{A}_{L}({\bar \nu}^{A}_{R})$ and 
$\nu^{A}_{R}({\bar \nu}^{A}_{L})$ correspond to the fermions of the usual and mirror space-times. Under such circumstances any parafermion of (\ref{5}) and (\ref{6}) appears in the field of 
emission as the unified system of the neutrino and antineutrino of the same space. 

At first sight, a formation of each of paraneutrinos
\begin{equation}
(\nu_{L}^{A}, \nu_{R}^{A}), \, \, \, \,
({\bar \nu}^{A}_{R}, {\bar \nu}^{A}_{L})
\label{46}
\end{equation}
relates usual and mirror spaces as a consequence of the ideas of a broken flavor symmetry law.
Such a connection, however, can exist only in the case when the gauge invariance is violated at 
the absence of the possibility of further restoration. Therefore, if it turns out that the nature itself unites flavor and gauge symmetries in a unified whole [15], then there is no doubt that 
a true number conservation law has never violated.

Another characteristic moment is the axial-vector structure [15] of internal space of the usual 
C-noninvariant particle. It states that whatever the main purposes each of earlier experiments [16,17] about a quasielastic axial-vector mass may serve as the first source of facts indicating 
to the availability in nature of C-odd types of particles of a kind of axial-vector mirror 
Minkowski space-time. 

As a consequence, neither of transitions (\ref{7}) and (\ref{8}) is forbidden by any conservation laws. They predict herewith those processes in which the left-handed long-lived truly neutral neutrino is converted into a right-handed short-lived axial-vector neutrino of the same flavor.
The right-handed neutrino of true neutrality interacts with matter in the same mode until it 
does not pass into a left-handed C-odd neutrino without change of his flavor. 

Thus, the existence of any transition of (\ref{7}) and (\ref{8}) expresses, in the case of a purely Coulomb part of current, the idea about that a truly neutral neutrino comes forward in the system as the source of an axial-vector mirror photon.

A fundamental role in nature of a gauge boson of this type and some above unnoted aspects of a new 
equation of the unified field theory of neutrinos of true neutrality call for special presentation.

\vspace{0.8cm}
\noindent
{\bf References}
\begin{enumerate}
\item
Sharafiddinov, R.S.: J. Phys. Nat. Sci. {\bf 4}, 1 (2013). arXiv:physics/0702233
\item
Sharafiddinov, R.S.: Bull. Am. Phys. Soc. {\bf 57}(16), KA.00069 (2012). 

arXiv:1004.0997[hep-ph]
\item
Majorana, E.: Il Nuovo Cimento {\bf 14}, 171 (1937)
\item
Sharafiddinov, R.S.: Phys. Essays {\bf 19}, 58 (2006). arXiv:hep-ph/0407262
\item
Zel’dovich, Ya.B.: Dokl. Akad. Nauk SSSR {\bf 91}, 1317 (1953)
\item
Konopinski, E.J., Mahmoud, H.: Phys. Rev. {\bf 92}, 1045 (1953)
\item
Sharafiddinov, R.S.: Bull Am. Phys. Soc. {\bf 58}(4), K2.00037 (2013). 

arXiv:1104.3623 [physics.gen-ph]
\item
Sharafiddinov, R.S.: Bull. Am. Phys. Soc. {\bf 58}(4), K2.00038 (2013). 

arXiv:1111.2089 [physics.gen-ph]
\item
Dirac, P.A.M.: Proc. Roy. Soc. Lond. {\bf A 117}, 610 (1928)
\item
Sharafiddinov, R.S.: Bull. Am. Phys. Soc. {\bf 59}(5), T1.00009 (2014). 
Spacetime Subst. {\bf 3}, 47 (2002). arXiv:physics/0305008
\item
Sharafiddinov, R.S.: Bull. Am. Phys. Soc. {\bf 59}(18), JP.00046 (2014)
\item
Yuldashev, B.S., Sharafiddinov, R.S.: Spacetime Subst. {\bf 5}, 137 (2004). 

arXiv:hep-ph/0510080
\item
Sharafiddinov, R.S.: Fizika {\bf B 16}, 1 (2007). arXiv:hep-ph/0512346
\item
Sharafiddinov, R.S.: Can. J. Phys. {\bf 93}, 1005 (2015). arXiv:1409.2397 [physics.gen-ph]
\item
Sharafiddinov, R.S.: Bull. Am. Phys. Soc. {\bf 59}(5), L1.00036 (2014)
\item
Kuzmin, K.S., Lyubushkin, V.V., Naumov, V.A.: Eur. Phys. J. {\bf C 54}, 517 (2008). 

arXiv:0712.4384 [hep-ph]
\item
The CMS Collaboration: Phys. Rev. {\bf D 74}, 052002 (2006). arXiv:hep-ex/0603034
\end{enumerate}
\end{document}